\documentclass[twocolumn,astrosymb,tighten]{aastex631}

\definecolor{mycolor}{RGB}{255,50,50}

\usepackage[utf8]{inputenc}
\usepackage{bm}
\usepackage{natbib}
\usepackage{graphicx}	
\usepackage{amsmath}	
\usepackage{amssymb}	
\usepackage{ulem}

\newcommand{\SFR}{\text{SFR}}

\newcommand{\Myr}{~\text{Myr}}

\newcommand{\Msun}{~\text{M}_\odot}
\newcommand{\pc}{~\text{pc}}
\newcommand{\kpc}{~\text{kpc}}
\newcommand{\vir}{{\text{vir}}}
\newcommand{\percent}{~\text{per~cent}}
\newcommand{\HHH}{\text{\textsc{H22}}}
\newcommand{\fracKMtoteps}{4.4}
\newcommand{\fracKMtotL}{5.4}
\newcommand{\avgE}{{\langle\varepsilon\rangle}}

\shorttitle{Starburst-induced gas-stars kinematic misalignment}
\shortauthors{Cenci et al.}

\begin{document}

\title{Starburst-induced gas-stars kinematic misalignment}

\author[0000-0002-0766-1704]{Elia Cenci}
\affiliation{Institute for Computational Science, University of Zurich, Winterthurerstrasse 190, Zurich CH-8057, Switzerland}
\correspondingauthor{Elia Cenci} \email{elia.cenci@uzh.ch}

\author[0000-0002-1109-1919]{Robert Feldmann}
\affiliation{Institute for Computational Science, University of Zurich, Winterthurerstrasse 190, Zurich CH-8057, Switzerland}

\author[0000-0001-6119-9883]{Jindra Gensior}
\affiliation{Institute for Computational Science, University of Zurich, Winterthurerstrasse 190, Zurich CH-8057, Switzerland}

\author[0000-0003-4298-5082]{James S. Bullock}
\affiliation{Department of Physics and Astronomy, University of California, Irvine, California 92697}

\author[0000-0002-3430-3232]{Jorge Moreno}
\affiliation{Department of Physics and Astronomy, Pomona College, Claremont, CA 91711, USA}
\affiliation{Centre for Computational Astrophysics, Flatiron Institute, 162 Fifth Avenue, New York, NY 10010, USA}

\author[0000-0002-6864-7762]{Luigi Bassini}
\affiliation{Institute for Computational Science, University of Zurich, Winterthurerstrasse 190, Zurich CH-8057, Switzerland}

\author[0000-0002-2930-9509]{Mauro Bernardini}
\affiliation{Institute for Computational Science, University of Zurich, Winterthurerstrasse 190, Zurich CH-8057, Switzerland}



\begin{abstract}

A kinematic misalignment of the stellar and gas components is a phenomenon observed in a significant fraction of galaxies. However, the underlying physical mechanisms are not well understood. A commonly proposed scenario for the formation of a misaligned component requires any pre-existing gas disc to be removed, via fly-bys or ejective feedback from an active galactic nucleus. In this Letter, we study the evolution of a Milky Way mass galaxy in the FIREbox cosmological volume that displays a thin, counter-rotating gas disc with respect to its stellar component at low redshift. In contrast to scenarios involving gas ejection, we find that pre-existing gas is mainly removed via the conversion into stars in a central starburst, triggered by a merging satellite galaxy. The newly-accreted, counter-rotating gas eventually settles into a kinematically misaligned disc. About $\fracKMtoteps\percent$ (8 out of 182) of FIREbox galaxies with stellar masses larger than $5\times 10^9\Msun$ at $z=0$ exhibit gas-star kinematic misalignment. In all cases, we identify central starburst-driven depletion as the main reason for the removal of the pre-existing co-rotating gas component, with no need for feedback from, e.g., a central active black hole. However, during the starburst, the gas is funneled towards the central regions, likely enhancing black hole activity. By comparing the fraction of misaligned discs between FIREbox and other simulations and observations, we conclude that this channel might have a non-negligible role in inducing kinematic misalignment in galaxies.

\end{abstract}



\section{Introduction} \label{sec:intro}
The kinematic properties of galaxies, such as the distribution and motion of stars and gas, provide important insights into their formation and evolution \citep[e.g.,][]{Somerville&Dave2015}. In general, the stellar component of galaxies is expected to inherit the kinematic properties of the gas out of which it forms, which are in turn set by those of their host dark matter halo \citep[e.g.,][]{Hoyle1951,Peebles1969}. Theoretical models also predict that the axisymmetric or triaxial potential of the stars exerts a torque on the gaseous component that leads to kinematic relaxation, i.e., the alignment of the average angular momenta of both components \citep{Tohline1982,Lake&Norman1983}. Thus, as a first approximation, both components (stars and gas) are expected to be substantially aligned. However, observations reveal that about $10\percent$ ($30\percent$) of late-type (early-type) galaxies exhibits either a gaseous or stellar component that is counter-rotating with respect to a co-spatial stellar component \citep[e.g.,][]{Galletta1987,Bertola1992,Rubin1992,Merrifield&Kuijken1994,Ciri1995,Bertola1996,Kuijken1996,Pizzella2004,Silchenko2009,Coccato2011,Pizzella2014,Pizzella2018,Silchenko2019,Proshina2020}.

Galaxies with misaligned components are also predicted by cosmological galaxy formation simulations \citep[][]{Khim2020,Koudmani2021,Duckworth2020b,Khoperskov2021} but with large variations, especially in the fraction of galaxies exhibiting gas-stars kinematic misalignment (KM). This fraction ranges between about 0.7 to $14\percent$, based on the misalignment angle between the angular momenta of gas and stars in central galaxies with stellar mass $M_\star\gtrsim 10^9\Msun$ at redshift $z=0$ \citep[e.g.,][]{Velliscig2015,Starkenburg2019,Casanueva2022}. Furthermore, galaxies with gas-stars counter-rotation likely exhibit also a kinematically misaligned stellar component and vice-versa, suggesting a common origin scenario \citep[][]{Khoperskov2021,Katkov2023}{}{}.

Numerical models also differ in the inferred mechanisms leading to KM. A number of theoretical studies argue that it is possible to achieve KM by either retrograde gas accretion \citep[][]{Algorry2014,Nedelchev2019,Khoperskov2021} or gas-rich minor mergers \citep{Thakar&Ryden1998,Bendo&Barnes2000,Bassett2017}. These scenarios are also well supported by observations of systems that show either evidence for accretion of counter-rotating material \citep[e.g.,][]{Chung2012,Osman&Bekki2017,Pizzella2018,Nedelchev2019,Bevacqua2022,Zhou2022,Katkov2023}{}{} or a recent/ongoing merger \citep[e.g.,][]{Thakar1997,Di_Matteo2008,Saburova2018,Katkov2023}{}{}. Gas-rich major mergers can also lead to the formation of counter-rotating components, especially when the merging, counter-rotating galaxies are co-planar \citep[e.g.,][]{Puerari&Pfenniger2001,Crocker2009}{}{}. 

In order for a galaxy to display KM, the new counter-rotating gas disc has to replace any extended pre-existing co-rotating gas disc. The latter must thus be removed, e.g., via ejective feedback from a active galactic nucleus \citep[AGN; e.g.,][]{Starkenburg2019}{}{} or from stripped by tidal effects from nearby satellites \citep[e.g.,][]{Starkenburg2019,Khoperskov2021}{}{}. Stellar feedback-driven outflows could also play a role in expelling the pre-existing gas \citep[e.g.,][]{Muratov2015,Pandya2021}{}{}, however, their role in relation to kinematic misalignment has received little attention. An alternative possibility to AGN/stellar-feedback or tidal stripping is that the pre-existing gas is depleted via star formation. The relative role of different mechanisms in effectively removing the pre-existing gas has not been explored. 

A correlation between AGN activity and the emergence of a counter-rotating gaseous component has been recently observed \citep[][]{Raimundo2023}{}{} and has also been predicted by cosmological and idealised simulations \citep[e.g.][]{Capelo&Dotti2017,Starkenburg2019,Duckworth2020a}. However, whether or not this AGN-KM correlation is indicative of AGN feedback expelling gas is still an open question \citep[see, e.g.,][]{Khoperskov2021}{}{}.


In this work, we consider a sample of galaxies from the FIREbox cosmological simulation \citep[][]{Feldmann2023} that exhibits KM at $z=0$. FIREbox simulates galaxy evolution with a high dynamic range and implements an accurate treatment of the interstellar medium physics by means of the FIRE-2 models \citep[][]{Hopkins2018}. FIREbox does not implement feedback from active black holes making it an ideal numerical experiment to explore the occurrence of KM in the absence of this feedback mechanism. In this Letter, we will mainly focus on a single galaxy of approximately Milky Way-mass at $z=0$. This galaxy displays a well-defined, counter-rotating gaseous disc, representing the most dramatic example of KM in FIREbox. We aim at understanding the physical processes responsible for removing any possible pre-existing co-rotating gas disc, promoting the formation of a misaligned component.

\section{FIREbox}\label{sec:FIREbox}
The galaxies analysed in this work are drawn from the FIREbox cosmological $\left(22.1~\text{cMpc}\right)^3$ volume \citep[][]{Feldmann2023}, that is part of the \textit{Feedback In Realistic Environments}\footnote{\url{https://FIRE.northwestern.edu}} (FIRE) project \citep{Hopkins2014,Hopkins2018}. The initial conditions refer to $z=120$ and were created with MUlti Scale Initial Conditions \citep[MUSIC;][]{Hahn&Abel2011}. The chosen cosmological parameters are consistent with Planck 2015 results \citep{Alves2016}: $\Omega_{\rm m}=0.3089$, $\Omega_\Lambda=1-\Omega_{\rm m}$, $\Omega_{\rm b}=0.0486$, $h=0.6774$, $\sigma_8=0.8159$, $n_{\rm s}=0.9667$. The transfer function is calculated with \textsc{camb}\footnote{\url{http://camb.info}} \citep{Lewis2000,Lewis2011}.

FIREbox is run with the meshless-finite-mass code \textsc{gizmo} \citep{Hopkins2015}, with FIRE-2 sub-grid physics \citep{Hopkins2018}. Gravity is calculated with a modified version of the parallelisation and tree gravity solver of GADGET-3 \citep{Springel2005b} that allows for adaptive force softening. Hydrodynamics is solved with the meshless-finite-mass method introduced in \citet{Hopkins2015}. 
The FIRE-2 model includes gas cooling down to 10 K, that naturally results in a multi-phase interstellar medium (ISM). Star formation occurs stochastically in dense ($n>300$ cm$^{-3}$), self-gravitating, Jeans-unstable, and self-shielded gas, with a $100\percent$ efficiency per free-fall time. Stellar feedback channels include energy, momentum, mass, and metal injections from type II and type Ia supernovae and stellar winds from OB and AGB stars. Radiative feedback models account for photo-ionisation and photo-electric heating and radiation pressure from young stars, using the Locally Extincted Background Radiation in Optically thin Networks (LEBRON) approximation \citep{Hopkins2012a}. Feedback from supermassive black holes is not included.

The FIRE-2 model has been validated in several studies on galaxy formation and evolution across wide ranges of stellar masses and numerical resolutions \citep{Wetzel2016,Hopkins2018,Ma2018a,Ma2018b}. FIREbox reproduces many of the observed galaxy properties \citep[see][]{Feldmann2023}{}{} and has been analysed in a number of recent studies \citep[e.g.,][]{Bernardini2022,Rohr2022,Moreno2022,Cenci2023,Gensior2023a,Gensior2023b}{}{}.

The analysis presented here is based on the fiducial FIREbox hydrodynamical simulation ($N_{\rm b}=1024^3$ and $N_{\rm DM}=1024^3$), with a mass resolution of $m_{\rm b}=6.3\times 10^4\Msun$ for baryons and $m_{\rm DM}=3.3\times 10^5\Msun$ for dark matter. The force resolution is $12\pc$ (physical, up to $z=9$; comoving for $z>9$) for stars and $80\pc$ for dark matter. The force softening of gas particles is adaptive (and related to their smoothing length) down to a minimum of $1.5\pc$ in the dense ISM.

\section{Definitions and selection}\label{sec:definitions}
To identify dark matter haloes we employ the AMIGA Halo Finder (AHF)\footnote{\url{http://popia.ft.uam.es/AHF/Download.htmlTable}} \citep[][]{Gill2004,Knollmann2009}. The halo radius $R_\vir$ is defined based on the virial overdensity criterion, so that the halo virial mass is:

\begin{equation}
 M_\vir = \frac{4\pi}{3} \Delta\left(z\right) \rho_{\rm m}\left(z\right) R_\vir^3 ~,
 \label{eqn:Mvir}
\end{equation}

where $\rho_{\rm m}\left(z\right)$ is the critical density at a given redshift, $\Delta\left(z\right)=\left( 18\pi^2 - 82 \Omega_\Lambda\left(z\right) - 39\left[\Omega_\Lambda\left(z\right)\right]^2 \right)/\Omega_{\rm m}\left(z\right)$ is the overdensity parameter, and $\Omega_\Lambda\left(z\right)$, $\Omega_{\rm m}\left(z\right)$ are the cosmological density parameters at redshift $z$ \citep[][]{Bryan&Norman1998}. We define the total galaxy radius to be $10\percent$ of the halo virial radius \citep[][]{Price2017}.

For each particle in a given galaxy, we define the circularity parameter as in \citet[][]{Abadi2003}:
\begin{equation}
 \varepsilon = \frac{j_z}{j_{\rm max}\left(E\right)}~,
 \label{eqn:eps}
\end{equation}
where $j_z$ is the projection of the specific angular momentum along the total specific angular momentum of the \textit{stars} in the galaxy (representing the $z$-axis in the reference frame we choose) and $j_{\rm max}\left(E\right)$ is the maximum specific angular momentum allowed on a circular orbit with the same specific binding energy $E$. Particles with $\varepsilon > 0$ ($\varepsilon < 0$) are on co-rotating (counter-rotating) orbits with respect to the average rotation axis of the stellar component. In the following, we will refer to the orientation of orbits with respect to the rotation axis of the stars. The average circularity, $\avgE$, computed as the mass-weighted arithmetic average of the circularity parameter of all the considered particles, gives a first estimate of the importance of the disc component and on the direction of rotation. For instance, $\avgE\sim 1$ ($\avgE\sim -1$) indicates the presence a prominent co-rotating (counter-rotating) disc component, while $\avgE\sim 0$ is related to the presence of either a dominant bulge component or equally important counter- and co-rotating components. A kinematic (co-rotating) disc component is commonly identified with the set of particles with a circularity parameter $\varepsilon \geq \varepsilon_{\rm th}$, typically with $\varepsilon_{\rm th}=0.7$ \citep[e.g.,][]{Aumer2013}. Similarly, we measure the contribution of the counter-rotating gas disc component as the mass fraction of gas particles in the galaxy with $\varepsilon\leq -\varepsilon_{\rm th}$.

We use the average circularity of all gas particles in the galaxy to determine if the galaxy shows KM. About $\fracKMtoteps\percent$ of all $z=0$ FIREbox galaxies with $M_\star\geq 5\times10^{9}\Msun$ have $\avgE\leq 0$. This fraction becomes $0.7\percent$ if we are more restrictive with our definition and demand $\avgE\leq -0.7$. Evaluating the degree of misalignment using the angle between the total angular momenta of the gas and stellar components gives similar results and it does not affect our conclusions. For comparison, about $\fracKMtotL\percent$ of $z=0$ FIREbox galaxies with $M_\star\geq 5\times10^{9}\Msun$ display a misalignment angle between the gas and stars angular momenta $\theta \geq$ 90 deg. This fraction reduces to about $1.7\percent$ for $\theta \geq$ 150 deg. 

In this work, we focus on studying a specific Milky Way mass galaxy at $z=0$, that represents the most dramatic case of kinematic misalignment in FIREbox. This galaxy displays a distribution of circularity parameters at $z=0$ that is consistent with having a well-defined, counter-rotating gas disc, with average circularity $\avgE\lesssim -0.9$. On the other hand, its stellar component has an $\avgE\sim 0.1$. 

\section{The two phases of the kinematic misalignment}

\begin{figure*}
 \centering
	\includegraphics[width=\textwidth]{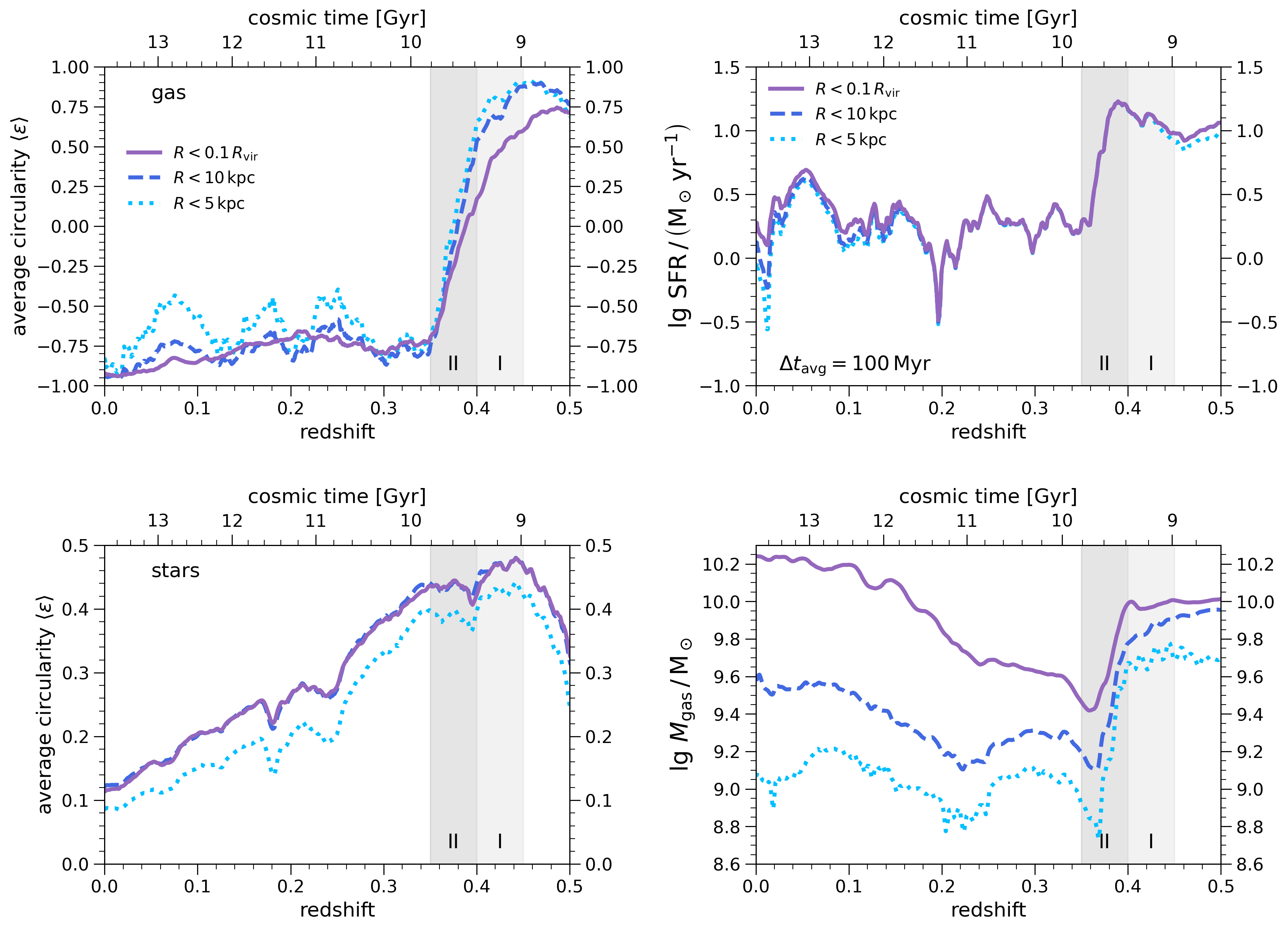}
 \caption{\textit{Left panels}: Temporal evolution of the average circularity parameter for the gas (top panel) and stars (bottom panel), computed with respect to the average angular momentum of the stellar component, within different apertures ($0.1\,R_{\rm vir}$, $10\kpc$, and $5\kpc$). The grey bands, at $z=0.45-0.4$ and $z=0.4-0.35$, highlight the time period during which the KM event approximately occurs. At redshift $z=0.5$ the gas disc is co-rotating with the stellar component. During $z\sim 0.45-0.35$, the gas then experiences a drastic KM over about $800\Myr$, settling into a well-defined, counter-rotating disc. The average circularity of stars increases from $0.3$ to $0.42$ in the first $\sim 300\Myr$ since $z=0.5$, but then progressively reduces to $\sim 0.1$ at $z=0$. \textit{Right panels}: Temporal evolution of the star formation rate ($\SFR$, averaged over $100\Myr$; top panel) and total gas mass ($M_{\rm gas}$; bottom panel), within different apertures ($0.1\,R_{\rm vir}$, $10\kpc$, and $5\kpc$). During the misalignment period (grey bands at $z=0.45-0.4$ and $z=0.4-0.35$), the $\SFR$ steadily increases by $\sim 0.2$ dex until $z\sim 0.38$ (for about $600\Myr$) to then drastically drop by a factor $\sim 1$ dex by $z\sim 0.35$, independent of the chosen aperture. At the same time, the total gas mass decreases slightly by $\sim 0.1-0.3$ dex until $z\sim 0.4$ to then swiftly drop by $\sim 0.6-1$ dex by $z=0.35$, depending on the chosen aperture.}
 \label{fig:eps_SFR_Mgas_vs_time}
\end{figure*}

Figure~\ref{fig:eps_SFR_Mgas_vs_time} shows the most dramatic example of KM from FIREbox, taking place in the galaxy that we will hereafter label as $\HHH$. At $z=0$, $\HHH$ has virial mass $M_{\rm vir}\sim 1.2\times 10^{12}\Msun$, stellar mass $M_\star\sim 6.3\times 10^{10}\Msun$, and total gas mass $M_{\rm gas}\sim 1.8\times 10^{10}\Msun$, within its size of $0.1\,R_{\rm vir}\sim 27.6\kpc$ at $z=0$. Close to the time of the KM event, $\HHH$ accretes gas mainly by tidal stripping from a gas-rich satellite galaxy. At $z\sim0.7$, the satellite was at its last apocenter before merging, at about $100\kpc$ from the center of $\HHH$. At that time, the satellite has stellar mass $M_\star\sim 10^{9}\Msun$ and total gas mass $M_{\rm gas}\sim 2\times 10^{9}\Msun$, both computed within a characteristic size of about $6\kpc$ (of the order twice its stellar half mass radius).

The left panels in Figure~\ref{fig:eps_SFR_Mgas_vs_time} show the temporal evolution of the average circularity, $\avgE$, for the gas (top panel) and stars (bottom panel) in $\HHH$. We compute $\avgE$ within $0.1\,R_{\rm vir}$ and within fixed apertures of $10\kpc$ and $5\kpc$. In the $\sim 800\Myr$ between redshift $z\sim0.35-0.45$, $\avgE$ of gas drops and switches signs from $\sim 0.8$ to $\sim -0.75$ (independent of the considered aperture). The average circularity then remains approximately constant at $\avgE\sim -0.75$ until $z=0.2$, after which it decreases to $\sim -0.9$ by $z=0$. We will hereafter refer to the transition period of time between $z\sim 0.35-0.45$ as the KM event.

The $\avgE$ of the stars within $0.1\,R_{\rm vir}$ and $10\kpc$ increases from $\sim 0.3$ to $\sim 0.42$, from redshift $z=0.5$ to $z\sim 0.45$, to then steadily decrease down to $\sim 0.1$ at $z=0$. These results are consistent with a steady counter-rotating gas disc is fueling the star formation in the galaxy until $z\sim 0$. 

The right panels in Figure~\ref{fig:eps_SFR_Mgas_vs_time} show the temporal evolution of the total star formation rate ($\SFR$, averaged over $100\Myr$) and total gas mass ($M_{\rm gas}$), within $0.1\,R_{\rm vir}$, $10\kpc$, and $5\kpc$. In the first $\sim 600\Myr$ during the misalignment event ($z\sim0.45-0.38$), the $\SFR$ steadily increases changing by $\sim 0.2$ dex, to then decrease by a factor $\sim 1$ dex in the following $\sim 200\Myr$ (until $z\sim 0.35$). The change in $\SFR$ over time is independent on the chosen aperture, implying that the bulk of the star formation is always occurring within $5\kpc$. Until $z=0$, the galaxy maintains an approximately constant $\SFR$, within $\sim 0.3$ dex. The total gas mass stays approximately constant (within $\sim 0.1$ dex) from $z\sim 0.45-0.4$, and then decreases by $\sim 0.6$ ($\sim 1$) dex until $z\sim 0.35$, within $0.1\,R_{\rm vir}$ ($5\kpc$). After $z\sim 0.35$, the total gas mass within $0.1\,R_{\rm vir}$ gradually increases by $\sim 0.8$ dex until $z=0$, while the total gas mass within smaller radii increases by only $\sim 0.5$ dex until $z=0$. 

To summarise, the misalignment event consists of mainly two steps. In the first phase, the $\SFR$ increases at constant total gas mass and the average gas (stellar) circularity decreases (increases). During the second phase the gas mass drops drastically, followed by a drop in $\SFR$ about $200\Myr$ later, and the average circularity of both the gas and stars decreases until complete KM is reached. 

\section{Depletion of the pre-existing gas and new accretion}
To study the fate of the co-rotating gas, prior to the emergence of the kinematically misaligned disc, we trace the properties of the gas particles constituting the pre-existing gas reservoir through time in the simulation.

\begin{figure*}
 \centering
 \includegraphics[width=\textwidth]{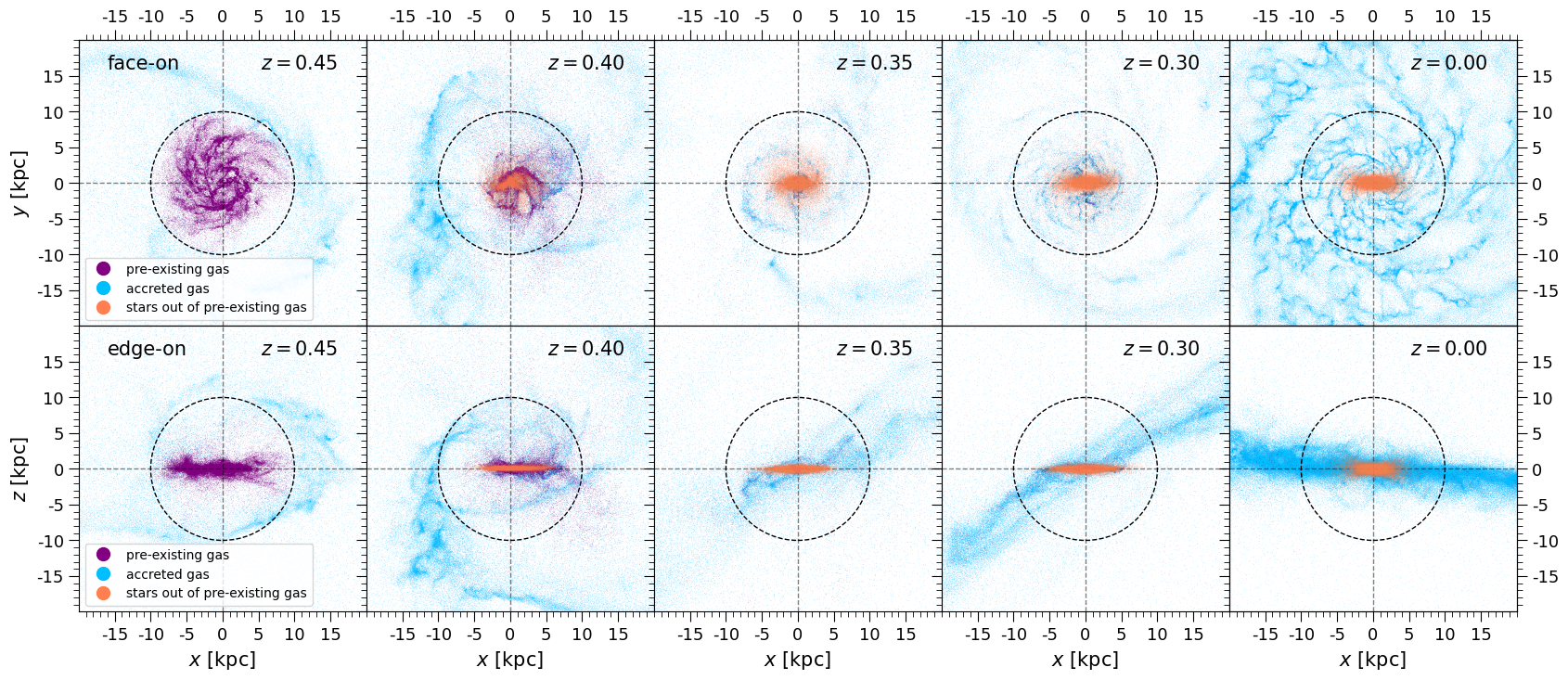}
 \caption{Face-on and edge-on projections (upper and lower panels, respectively) of $\HHH$ in four snapshots around the period of KM and at $z=0$. Gas particles are color-coded according to whether they are within or outside the $10\kpc$ from the centre of the galaxy at $z=0.45$ (purple and blue points, respectively), when the pre-existing gas disc is still in place. The projections are rotated in the $\left(x,y\right)$-plane such that the $x$-axis is aligned with the principal inertia axis associated with the largest moment. The pre-existing gas appears to be depleted from the galaxy by $z\sim 0.35$ and the counter-rotating gas accreted from the merging satellite settles into a thin, misaligned gas disc by $z\sim 0.3$. Stars formed out of the pre-existing gas (orange points) remain within the galaxy until $z=0.3$, forming a central, thin, bar-like structure. At $z=0$, the stars formed from the pre-existing baryons are still found in the central regions of $\HHH$.}
 \label{fig:gas_maps}
\end{figure*}

Figure~\ref{fig:gas_maps} shows the evolution of the total gas content in $\HHH$, during $z=0.3-0.45$ when the KM occurs. Particles are color-coded based on whether they constitute the pre-existing gas reservoir, i.e., whether they are within $10\kpc$ at $z=0.45$. By $z\sim 0.35$ the gas content of $\HHH$ is almost entirely replaced by the gas that has been inflowing onto the galaxy from a merging satellite. Furthermore, we show that most of the stars formed out of the pre-existing gas reside within the galaxy at $z=0.3$ and turned into stars, rather than expelled/stripped from the galaxy. At $z=0$, the stars formed out of the pre-existing gas disc are found in a thicker stellar structure in the innermost $2\kpc$ of $\HHH$ \citep[see, e.g.,][and references therein]{Yu2021,Yu2023}{}{}, while most of the new stars at $z=0$ are forming in the more extended counter-rotating gaseous disc.

\begin{figure*}
 \centering
	\includegraphics[width=\textwidth]{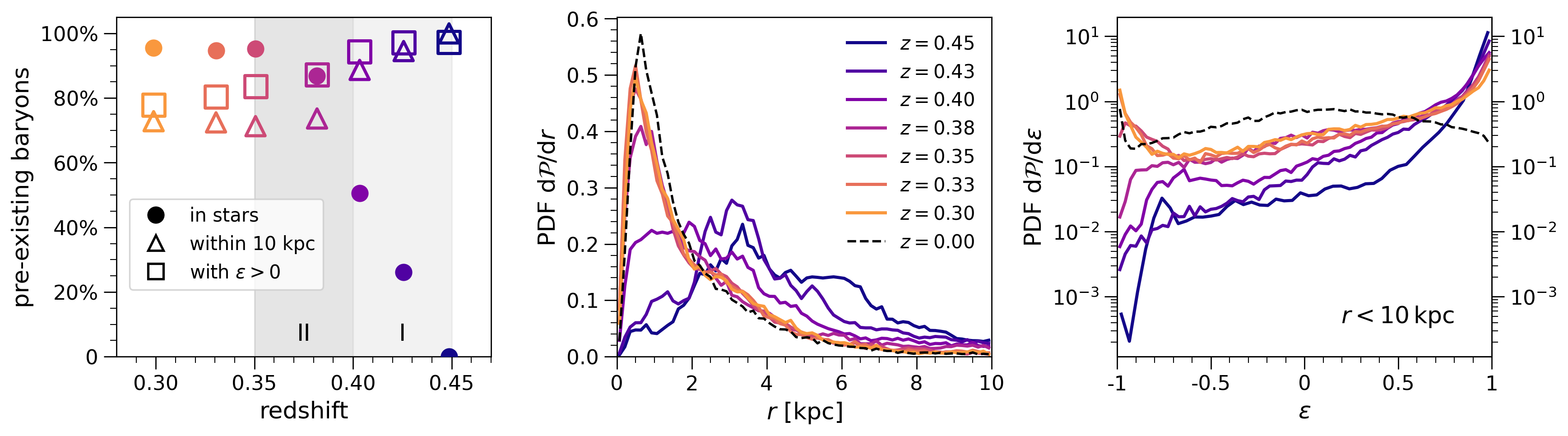}
 \caption{Evolution of the baryons forming the pre-existing gas disc at $z=0.45$ (i.e., the pre-existing baryons). \textit{Left panel}: evolution of the fraction of pre-existing baryons that formed stars (circles), that is still within the initial aperture of $10\kpc$ (triangles), and that has a circularity parameter $\varepsilon>0$ (i.e., on co-rotating orbits with respect to the stellar component of the galaxy). \textit{Middle panel}: mass-weighted probability density function (PDF) of the galactocentric radii of the pre-existing baryons. \textit{Right panel}: mass-weighted PDF of the circularity parameter of pre-existing baryons with respect to stars, within $10\kpc$. The color-code refers to redshift. About $70\percent$ of the pre-existing baryons is still within the galaxy by $z=0.3$ and concentrates within $2-5\kpc$. About $95\percent$ of the pre-existing baryons is converted into stars already by $z\sim 0.35$. About $80\percent$ of the pre-existing baryons retain a circularity parameter $\varepsilon>0$.}
 \label{fig:traced_gas_particles_R10kpc}
\end{figure*}

Figure~\ref{fig:traced_gas_particles_R10kpc} shows how the baryons constituting the pre-existing gas reservoir within $10\kpc$ at $z=0.45$ (i.e., the pre-existing baryons) progressively move to smaller galactocentric radii and form stars. About $80\percent$ of the the pre-existing baryons in $\HHH$ are still found within the galaxy at $z\sim 0.3$, especially within $\sim 2-5\kpc$. About $95\percent$ of the pre-existing baryons turn into stars already by $z=0.35$. Furthermore, approximately $80\percent$ of them retain a circularity parameter $\varepsilon>0$, implying that they mainly contribute to the co-rotating component, although not necessarily to the disc. Furthermore, we find that about $45\percent$ ($10\percent$) of pre-existing baryons have $\varepsilon > 0.7$ ($\varepsilon < -0.7$) at $z=0.3$ and therefore contribute to the co-rotating (counter-rotating) stellar disc. By $z=0$, the pre-existing baryons are still found in $\HHH$ and have a more bulge-like kinematic morphology, based on the distribution of their circularities, with only $20\percent$ of them contributing to the (either co- or counter-rotating) stellar disc component.

\section{Summary and discussion}\label{sec:discussion}
In this Letter, we studied the origin of the kinematic misalignment between the gas and stellar components (KM) of $z=0$ galaxies in the FIREbox cosmological volume \citep[][]{Feldmann2023}{}{}. Specifically, we aimed at understanding the mechanisms causing the depletion of the pre-existing gas reservoir and thus allowing for the formation of a stable, counter-rotating gas disc.

Our main finding is that the pre-existing gas disc that was co-rotating with respect to stars at $z=0.45$ is almost entirely ($\sim 95\percent$) converted into stars by $z=0.3$ (in about $800\Myr$; see Figure~\ref{fig:traced_gas_particles_R10kpc}, left and middle panels), rather than being ejected. This compaction-driven starburst \citep[see, e.g.,][]{Dekel&Burkert2014,Zolotov2015,Tacchella2016a,Cenci2023} is triggered by instabilities induced by a nearby gas-rich merging satellite.

In this case study, KM occurs in essentially two phases. In the first phase, torques arising from the interaction destabilize the co-rotating gas disc against global gravitational collapse, driving a gas compaction event and increasing the star-formation rate \citep[see, e.g.,][]{Barnes&Hernquist1991,Mihos&Hernquist1994,Mihos&Hernquist1996,Hopkins2006,Di_Matteo2007,Cox2008,Renaud2014,Moreno2015,Hopkins2018,Renaud2019,Moreno2021,SegoviaOtero2022,Cenci2023}{}{}. As a result of co-rotating gas being converted into stars, the average circularity of the stellar component increases while the average circularity for the gas starts decreasing. In the second phase, gas is rapidly depleted and converted into stars and the newly accreted gas from the merging satellite gradually forms a star-forming counter-rotating disc with respect to stars. Consequently, the average circularity of both the stellar and gas components decreases.

About $80\percent$ of the stars formed out of the pre-existing gas retain a co-rotating orbital configuration by the end of the KM period. The newly accreted counter-rotating gas affects the orbits of the star-forming, pre-existing gas, increasing the fraction of pre-existing baryons on counter-rotating orbits, especially in the second phase of KM.

A starburst-induced KM may occur repeatedly during the lifetime of galaxies and its impact is possibly imprinted in the distributions of ages and metallicities for stars in the co-rotating and counter-rotating components, as reported by observations \citep[e.g.,][]{Pizzella2014,Bassett2017,Nedelchev2019,Katkov2023,Zinchenko2023}. This naturally depends on the origin of the newly accreted gas and it is therefore not trivial to pinpoint when KM happened. Compared to a KM induced by AGN feedback, the drop in $\SFR$ and total gas mass in the galaxy happens on a longer time-scale (of the order of the gas depletion time) in the case of a starburst-induced KM. Furthermore, we showed that most of the stars formed during KM would reside in the central region of the galaxy, as opposed to what expected if the pre-existing gas was expelled from the galaxy by AGN feedback. High spectral and spatial resolution data from, e.g., MUSE \citep[Multi Unit Spectroscopic Explorer;][]{Bacon2010} and JWST \citep[James Webb Space Telescope;][]{Gardner2006} will enable a more complete analysis of the kinematics, metallicities, and ages of stars in galaxies with a counter-rotating component, helping to shed light on the exact mechanism that lead to KM.

Our starburst scenario promotes central gas inflow, potentially fuelling the central black hole. The AGN-KM connection might thus arise from central gas compaction events, inducing both the AGN activity and removal of the pre-existing gas disc. This picture is consistent with recent observations of a large fraction of galaxies exhibiting both kinematically misaligned components and active galactic nuclei \citep[AGN; e.g.,][]{Raimundo2023}{}{}. It also aligns with theoretical work arguing that gas removal by AGN feedback is sub-dominant compared to other processes such as, e.g., tidal stripping \citep[e.g.,][]{Khoperskov2021}{}{}.

About $\fracKMtoteps\percent$ of $z=0$ galaxies in FIREbox displays gas-star kinematic misalignment (KM), comparable to those reported in previous numerical works \citep[e.g.,][]{Velliscig2015,Starkenburg2019,Casanueva2022}. Therefore, the channel we propose plays a potentially important role in driving KM.

We showed that merger-induced starbursts may be a novel mechanism to promote gas-stars kinematic misalignment in galaxies. However, its imprint on the galaxy properties might not be easily disentangled from that of scenarios involving black hole physics. Further work with larger samples of galaxies displaying KM is thus required to quantify the relative importance of this channel and its implications for galaxy evolution.

\section*{Data Availability}
The data underlying this article are available on reasonable request to the corresponding author. A public version of the \textsc{gizmo} code is available at \url{http://www.tapir.caltech.edu/~phopkins/Site/GIZMO.html}. FIRE data releases are publicly available at \url{http://flathub.flatironinstitute.org/fire}.

\begin{acknowledgements}
    EC thanks Papa IV for His countless blessings. EC, RF, JG, LB, MB acknowledge financial support from the Swiss National Science Foundation (grant no PP00P2$\_$194814). EC, RF, LB, MB acknowledge financial support from the Swiss National Science Foundation (grant no 200021$\_$188552). JG gratefully acknowledges financial support from the Swiss National Science Foundation (grant no CRSII5$\_$193826). JM is funded by the Hirsch Foundation. This work made use of infrastructure services provided by S3IT (\url{www.s3it.uzh.ch}), the Service and Support for Science IT team at the University of Zurich. All plots were created with the \textsc{matplotlib} library for visualization with Python \citep{Hunter2007}.
\end{acknowledgements}



\bibliography{main}{}

\begin{thebibliography}{}
\expandafter\ifx\csname natexlab\endcsname\relax\def\natexlab#1{#1}\fi
\providecommand{\url}[1]{\href{#1}{#1}}
\providecommand{\dodoi}[1]{doi:~\href{http://doi.org/#1}{\nolinkurl{#1}}}
\providecommand{\doeprint}[1]{\href{http://ascl.net/#1}{\nolinkurl{http://ascl.net/#1}}}
\providecommand{\doarXiv}[1]{\href{https://arxiv.org/abs/#1}{\nolinkurl{https://arxiv.org/abs/#1}}}

\bibitem[{{Abadi} {et~al.}(2003){Abadi}, {Navarro}, {Steinmetz}, \&
  {Eke}}]{Abadi2003}
{Abadi}, M.~G., {Navarro}, J.~F., {Steinmetz}, M., \& {Eke}, V.~R. 2003, \apj,
  597, 21, \dodoi{10.1086/378316}

\bibitem[{{Algorry} {et~al.}(2014){Algorry}, {Navarro}, {Abadi}, {Sales},
  {Steinmetz}, \& {Piontek}}]{Algorry2014}
{Algorry}, D.~G., {Navarro}, J.~F., {Abadi}, M.~G., {et~al.} 2014, \mnras, 437,
  3596, \dodoi{10.1093/mnras/stt2154}

\bibitem[{Alves {et~al.}(2016)Alves, Combes, Ferrara, Forveille, \&
  Shore}]{Alves2016}
Alves, J., Combes, F., Ferrara, A., Forveille, T., \& Shore, S. 2016, Astronomy
  and Astrophysics, 594, \dodoi{10.1051/0004-6361/201629543}

\bibitem[{{Aumer} {et~al.}(2013){Aumer}, {White}, {Naab}, \&
  {Scannapieco}}]{Aumer2013}
{Aumer}, M., {White}, S. D.~M., {Naab}, T., \& {Scannapieco}, C. 2013, \mnras,
  434, 3142, \dodoi{10.1093/mnras/stt1230}

\bibitem[{{Bacon} {et~al.}(2010){Bacon}, {Accardo}, {Adjali}, {Anwand},
  {Bauer}, {Biswas}, {Blaizot}, {Boudon}, {Brau-Nogue}, {Brinchmann},
  {Caillier}, {Capoani}, {Carollo}, {Contini}, {Couderc}, {Daguis{\'e}},
  {Deiries}, {Delabre}, {Dreizler}, {Dubois}, {Dupieux}, {Dupuy}, {Emsellem},
  {Fechner}, {Fleischmann}, {Fran{\c{c}}ois}, {Gallou}, {Gharsa}, {Glindemann},
  {Gojak}, {Guiderdoni}, {Hansali}, {Hahn}, {Jarno}, {Kelz}, {Koehler},
  {Kosmalski}, {Laurent}, {Le Floch}, {Lilly}, {Lizon}, {Loupias}, {Manescau},
  {Monstein}, {Nicklas}, {Olaya}, {Pares}, {Pasquini}, {P{\'e}contal-Rousset},
  {Pell{\'o}}, {Petit}, {Popow}, {Reiss}, {Remillieux}, {Renault}, {Roth},
  {Rupprecht}, {Serre}, {Schaye}, {Soucail}, {Steinmetz}, {Streicher}, {Stuik},
  {Valentin}, {Vernet}, {Weilbacher}, {Wisotzki}, \& {Yerle}}]{Bacon2010}
{Bacon}, R., {Accardo}, M., {Adjali}, L., {et~al.} 2010, in Society of
  Photo-Optical Instrumentation Engineers (SPIE) Conference Series, Vol. 7735,
  Ground-based and Airborne Instrumentation for Astronomy III, ed. I.~S.
  {McLean}, S.~K. {Ramsay}, \& H.~{Takami}, 773508

\bibitem[{{Barnes} \& {Hernquist}(1991)}]{Barnes&Hernquist1991}
{Barnes}, J.~E., \& {Hernquist}, L.~E. 1991, \apjl, 370, L65,
  \dodoi{10.1086/185978}

\bibitem[{{Bassett} {et~al.}(2017){Bassett}, {Bekki}, {Cortese}, \&
  {Couch}}]{Bassett2017}
{Bassett}, R., {Bekki}, K., {Cortese}, L., \& {Couch}, W. 2017, \mnras, 471,
  1892, \dodoi{10.1093/mnras/stx958}

\bibitem[{{Bendo} \& {Barnes}(2000)}]{Bendo&Barnes2000}
{Bendo}, G.~J., \& {Barnes}, J.~E. 2000, \mnras, 316, 315,
  \dodoi{10.1046/j.1365-8711.2000.03475.x}

\bibitem[{{Bernardini} {et~al.}(2022){Bernardini}, {Feldmann},
  {Angl{\'e}s-Alc{\'a}zar}, {Boylan-Kolchin}, {Bullock}, {Mayer}, \&
  {Stadel}}]{Bernardini2022}
{Bernardini}, M., {Feldmann}, R., {Angl{\'e}s-Alc{\'a}zar}, D., {et~al.} 2022,
  \mnras, 509, 1323, \dodoi{10.1093/mnras/stab3088}

\bibitem[{{Bertola} {et~al.}(1992){Bertola}, {Buson}, \&
  {Zeilinger}}]{Bertola1992}
{Bertola}, F., {Buson}, L.~M., \& {Zeilinger}, W.~W. 1992, \apjl, 401, L79,
  \dodoi{10.1086/186675}

\bibitem[{{Bertola} {et~al.}(1996){Bertola}, {Cinzano}, {Corsini}, {Pizzella},
  {Persic}, \& {Salucci}}]{Bertola1996}
{Bertola}, F., {Cinzano}, P., {Corsini}, E.~M., {et~al.} 1996, \apjl, 458, L67,
  \dodoi{10.1086/309924}

\bibitem[{{Bevacqua} {et~al.}(2022){Bevacqua}, {Cappellari}, \&
  {Pellegrini}}]{Bevacqua2022}
{Bevacqua}, D., {Cappellari}, M., \& {Pellegrini}, S. 2022, \mnras, 511, 139,
  \dodoi{10.1093/mnras/stab3732}

\bibitem[{Bryan \& Norman(1998)}]{Bryan&Norman1998}
Bryan, G.~L., \& Norman, M.~L. 1998, The Astrophysical Journal, 495, 80,
  \dodoi{10.1086/305262}

\bibitem[{{Capelo} \& {Dotti}(2017)}]{Capelo&Dotti2017}
{Capelo}, P.~R., \& {Dotti}, M. 2017, \mnras, 465, 2643,
  \dodoi{10.1093/mnras/stw2872}

\bibitem[{{Casanueva} {et~al.}(2022){Casanueva}, {Lagos}, {Padilla}, \&
  {Davison}}]{Casanueva2022}
{Casanueva}, C.~I., {Lagos}, C. d.~P., {Padilla}, N.~D., \& {Davison}, T.~A.
  2022, \mnras, 514, 2031, \dodoi{10.1093/mnras/stac523}

\bibitem[{{Cenci} {et~al.}(2023){Cenci}, {Feldmann}, {Gensior}, {Moreno},
  {Bassini}, \& {Bernardini}}]{Cenci2023}
{Cenci}, E., {Feldmann}, R., {Gensior}, J., {et~al.} 2023, \mnras,
  \dodoi{10.1093/mnras/stad3709}

\bibitem[{{Chung} {et~al.}(2012){Chung}, {Bureau}, {van Gorkom}, \&
  {Koribalski}}]{Chung2012}
{Chung}, A., {Bureau}, M., {van Gorkom}, J.~H., \& {Koribalski}, B. 2012,
  \mnras, 422, 1083, \dodoi{10.1111/j.1365-2966.2012.20679.x}

\bibitem[{{Ciri} {et~al.}(1995){Ciri}, {Bettoni}, \& {Galletta}}]{Ciri1995}
{Ciri}, R., {Bettoni}, D., \& {Galletta}, G. 1995, \nat, 375, 661,
  \dodoi{10.1038/375661a0}

\bibitem[{{Coccato} {et~al.}(2011){Coccato}, {Morelli}, {Corsini}, {Buson},
  {Pizzella}, {Vergani}, \& {Bertola}}]{Coccato2011}
{Coccato}, L., {Morelli}, L., {Corsini}, E.~M., {et~al.} 2011, \mnras, 412,
  L113, \dodoi{10.1111/j.1745-3933.2011.01016.x}

\bibitem[{{Cox} {et~al.}(2008){Cox}, {Jonsson}, {Somerville}, {Primack}, \&
  {Dekel}}]{Cox2008}
{Cox}, T.~J., {Jonsson}, P., {Somerville}, R.~S., {Primack}, J.~R., \& {Dekel},
  A. 2008, \mnras, 384, 386, \dodoi{10.1111/j.1365-2966.2007.12730.x}

\bibitem[{{Crocker} {et~al.}(2009){Crocker}, {Jeong}, {Komugi}, {Combes},
  {Bureau}, {Young}, \& {Yi}}]{Crocker2009}
{Crocker}, A.~F., {Jeong}, H., {Komugi}, S., {et~al.} 2009, \mnras, 393, 1255,
  \dodoi{10.1111/j.1365-2966.2008.14295.x}

\bibitem[{{Dekel} \& {Burkert}(2014)}]{Dekel&Burkert2014}
{Dekel}, A., \& {Burkert}, A. 2014, \mnras, 438, 1870,
  \dodoi{10.1093/mnras/stt2331}

\bibitem[{{Di Matteo} {et~al.}(2008){Di Matteo}, {Bournaud}, {Martig},
  {Combes}, {Melchior}, \& {Semelin}}]{Di_Matteo2008}
{Di Matteo}, P., {Bournaud}, F., {Martig}, M., {et~al.} 2008, \aap, 492, 31,
  \dodoi{10.1051/0004-6361:200809480}

\bibitem[{{Di Matteo} {et~al.}(2007){Di Matteo}, {Combes}, {Melchior}, \&
  {Semelin}}]{Di_Matteo2007}
{Di Matteo}, P., {Combes}, F., {Melchior}, A.~L., \& {Semelin}, B. 2007, \aap,
  468, 61, \dodoi{10.1051/0004-6361:20066959}

\bibitem[{{Duckworth} {et~al.}(2020{\natexlab{a}}){Duckworth}, {Starkenburg},
  {Genel}, {Davis}, {Habouzit}, {Kraljic}, \& {Tojeiro}}]{Duckworth2020b}
{Duckworth}, C., {Starkenburg}, T.~K., {Genel}, S., {et~al.}
  2020{\natexlab{a}}, \mnras, 495, 4542, \dodoi{10.1093/mnras/staa1494}

\bibitem[{{Duckworth} {et~al.}(2020{\natexlab{b}}){Duckworth}, {Tojeiro}, \&
  {Kraljic}}]{Duckworth2020a}
{Duckworth}, C., {Tojeiro}, R., \& {Kraljic}, K. 2020{\natexlab{b}}, \mnras,
  492, 1869, \dodoi{10.1093/mnras/stz3575}

\bibitem[{{Feldmann} {et~al.}(2023){Feldmann}, {Quataert},
  {Faucher-Gigu{\`e}re}, {Hopkins}, {{\c{C}}atmabacak}, {Kere{\v{s}}},
  {Bassini}, {Bernardini}, {Bullock}, {Cenci}, {Gensior}, {Liang}, {Moreno}, \&
  {Wetzel}}]{Feldmann2023}
{Feldmann}, R., {Quataert}, E., {Faucher-Gigu{\`e}re}, C.-A., {et~al.} 2023,
  \mnras, 522, 3831, \dodoi{10.1093/mnras/stad1205}

\bibitem[{{Galletta}(1987)}]{Galletta1987}
{Galletta}, G. 1987, \apj, 318, 531, \dodoi{10.1086/165389}

\bibitem[{{Gardner} {et~al.}(2006){Gardner}, {Mather}, {Clampin}, {Doyon},
  {Greenhouse}, {Hammel}, {Hutchings}, {Jakobsen}, {Lilly}, {Long}, {Lunine},
  {McCaughrean}, {Mountain}, {Nella}, {Rieke}, {Rieke}, {Rix}, {Smith},
  {Sonneborn}, {Stiavelli}, {Stockman}, {Windhorst}, \& {Wright}}]{Gardner2006}
{Gardner}, J.~P., {Mather}, J.~C., {Clampin}, M., {et~al.} 2006, \ssr, 123,
  485, \dodoi{10.1007/s11214-006-8315-7}

\bibitem[{{Gensior} {et~al.}(2023{\natexlab{a}}){Gensior}, {Feldmann}, {Mayer},
  {Wetzel}, {Hopkins}, \& {Faucher-Gigu{\`e}re}}]{Gensior2023a}
{Gensior}, J., {Feldmann}, R., {Mayer}, L., {et~al.} 2023{\natexlab{a}},
  \mnras, 518, L63, \dodoi{10.1093/mnrasl/slac138}

\bibitem[{{Gensior} {et~al.}(2023{\natexlab{b}}){Gensior}, {Feldmann},
  {Reina-Campos}, {Trujillo-Gomez}, {Mayer}, {Keller}, {Wetzel}, {Kruijssen},
  {Hopkins}, \& {Moreno}}]{Gensior2023b}
{Gensior}, J., {Feldmann}, R., {Reina-Campos}, M., {et~al.} 2023{\natexlab{b}},
  arXiv e-prints, arXiv:2310.01482, \dodoi{10.48550/arXiv.2310.01482}

\bibitem[{{Gill} {et~al.}(2004){Gill}, {Knebe}, \& {Gibson}}]{Gill2004}
{Gill}, S. P.~D., {Knebe}, A., \& {Gibson}, B.~K. 2004, \mnras, 351, 399,
  \dodoi{10.1111/j.1365-2966.2004.07786.x}

\bibitem[{Hahn \& Abel(2011)}]{Hahn&Abel2011}
Hahn, O., \& Abel, T. 2011, Monthly Notices of the Royal Astronomical Society,
  415, 2101, \dodoi{10.1111/j.1365-2966.2011.18820.x}

\bibitem[{Hopkins(2015)}]{Hopkins2015}
Hopkins, P.~F. 2015, Monthly Notices of the Royal Astronomical Society,
  \dodoi{10.1093/mnras/stv195}

\bibitem[{{Hopkins} {et~al.}(2006){Hopkins}, {Hernquist}, {Cox}, {Di Matteo},
  {Robertson}, \& {Springel}}]{Hopkins2006}
{Hopkins}, P.~F., {Hernquist}, L., {Cox}, T.~J., {et~al.} 2006, \apjs, 163, 1,
  \dodoi{10.1086/499298}

\bibitem[{Hopkins {et~al.}(2014)Hopkins, Kere{\v{s}}, O{\~{n}}orbe,
  Faucher-Gigu{\`{e}}re, Quataert, Murray, \& Bullock}]{Hopkins2014}
Hopkins, P.~F., Kere{\v{s}}, D., O{\~{n}}orbe, J., {et~al.} 2014, Monthly
  Notices of the Royal Astronomical Society, \dodoi{10.1093/mnras/stu1738}

\bibitem[{Hopkins {et~al.}(2012)Hopkins, Quataert, \& Murray}]{Hopkins2012a}
Hopkins, P.~F., Quataert, E., \& Murray, N. 2012, Monthly Notices of the Royal
  Astronomical Society, 421, 3488, \dodoi{10.1111/J.1365-2966.2012.20578.X}

\bibitem[{Hopkins {et~al.}(2018)Hopkins, Wetzel, Kere{\v{s}},
  Faucher-Gigu{\`{e}}re, Quataert, Boylan-Kolchin, Murray, Hayward,
  Garrison-Kimmel, Hummels, Feldmann, Torrey, Ma, Angl{\'{e}}s-Alc{\'{a}}zar,
  Su, Orr, Schmitz, Escala, Sanderson, Grudi{\'{c}}, Hafen, Kim, Fitts,
  Bullock, Wheeler, Chan, Elbert, \& Narayanan}]{Hopkins2018}
Hopkins, P.~F., Wetzel, A., Kere{\v{s}}, D., {et~al.} 2018, Monthly Notices of
  the Royal Astronomical Society, 480, 800, \dodoi{10.1093/mnras/sty1690}

\bibitem[{{Hoyle}(1951)}]{Hoyle1951}
{Hoyle}, F. 1951, in Problems of Cosmical Aerodynamics, 195

\bibitem[{{Hunter}(2007)}]{Hunter2007}
{Hunter}, J.~D. 2007, Computing in Science and Engineering, 9, 90,
  \dodoi{10.1109/MCSE.2007.55}

\bibitem[{{Katkov} {et~al.}(2023){Katkov}, {Gasymov}, {Kniazev}, {Gelfand},
  {Rubtsov}, {Chilingarian}, \& {Sil'chenko}}]{Katkov2023}
{Katkov}, I., {Gasymov}, D., {Kniazev}, A., {et~al.} 2023, arXiv e-prints,
  arXiv:2305.01719, \dodoi{10.48550/arXiv.2305.01719}

\bibitem[{{Khim} {et~al.}(2020){Khim}, {Yi}, {Dubois}, {Bryant}, {Pichon},
  {Croom}, {Bland-Hawthorn}, {Brough}, {Choi}, {Devriendt}, {Groves}, {Owers},
  {Richards}, {van de Sande}, \& {Sweet}}]{Khim2020}
{Khim}, D.~J., {Yi}, S.~K., {Dubois}, Y., {et~al.} 2020, \apj, 894, 106,
  \dodoi{10.3847/1538-4357/ab88a9}

\bibitem[{{Khoperskov} {et~al.}(2021){Khoperskov}, {Zinchenko}, {Avramov},
  {Khrapov}, {Berczik}, {Saburova}, {Ishchenko}, {Khoperskov}, {Pulsoni},
  {Venichenko}, {Bizyaev}, \& {Moiseev}}]{Khoperskov2021}
{Khoperskov}, S., {Zinchenko}, I., {Avramov}, B., {et~al.} 2021, \mnras, 500,
  3870, \dodoi{10.1093/mnras/staa3330}

\bibitem[{Knollmann \& Knebe(2009)}]{Knollmann2009}
Knollmann, S.~R., \& Knebe, A. 2009, Astrophysical Journal, Supplement Series,
  182, 608, \dodoi{10.1088/0067-0049/182/2/608}

\bibitem[{{Koudmani} {et~al.}(2021){Koudmani}, {Henden}, \&
  {Sijacki}}]{Koudmani2021}
{Koudmani}, S., {Henden}, N.~A., \& {Sijacki}, D. 2021, \mnras, 503, 3568,
  \dodoi{10.1093/mnras/stab677}

\bibitem[{{Kuijken} {et~al.}(1996){Kuijken}, {Fisher}, \&
  {Merrifield}}]{Kuijken1996}
{Kuijken}, K., {Fisher}, D., \& {Merrifield}, M.~R. 1996, \mnras, 283, 543,
  \dodoi{10.1093/mnras/283.2.543}

\bibitem[{{Lake} \& {Norman}(1983)}]{Lake&Norman1983}
{Lake}, G., \& {Norman}, C. 1983, \apj, 270, 51, \dodoi{10.1086/161097}

\bibitem[{Lewis {et~al.}(2011)Lewis, Challinor, \& Hanson}]{Lewis2011}
Lewis, A., Challinor, A., \& Hanson, D. 2011, Journal of Cosmology and
  Astroparticle Physics, 2011, \dodoi{10.1088/1475-7516/2011/03/018}

\bibitem[{Lewis {et~al.}(2000)Lewis, Challinor, \& Lasenby}]{Lewis2000}
Lewis, A., Challinor, A., \& Lasenby, A. 2000, The Astrophysical Journal, 538,
  473, \dodoi{10.1086/309179}

\bibitem[{Ma {et~al.}(2018{\natexlab{a}})Ma, Hopkins, Boylan-Kolchin,
  Faucher-Gigu{\`{e}}re, Quataert, Feldmann, Garrison-Kimmel, Hayward,
  Kere{\v{s}}, \& Wetzel}]{Ma2018a}
Ma, X., Hopkins, P.~F., Boylan-Kolchin, M., {et~al.} 2018{\natexlab{a}},
  Monthly Notices of the Royal Astronomical Society, 477, 219,
  \dodoi{10.1093/mnras/sty684}

\bibitem[{Ma {et~al.}(2018{\natexlab{b}})Ma, Hopkins, Garrison-Kimmel,
  Faucher-Gigu{\'{e}}re, Quataert, Boylan-Kolchin, Hayward, Feldmann, \&
  Kere{\v{s}}}]{Ma2018b}
Ma, X., Hopkins, P.~F., Garrison-Kimmel, S., {et~al.} 2018{\natexlab{b}},
  Monthly Notices of the Royal Astronomical Society, 478, 1694,
  \dodoi{10.1093/mnras/sty1024}

\bibitem[{{Merrifield} \& {Kuijken}(1994)}]{Merrifield&Kuijken1994}
{Merrifield}, M.~R., \& {Kuijken}, K. 1994, \apj, 432, 575,
  \dodoi{10.1086/174596}

\bibitem[{{Mihos} \& {Hernquist}(1994)}]{Mihos&Hernquist1994}
{Mihos}, J.~C., \& {Hernquist}, L. 1994, \apjl, 431, L9, \dodoi{10.1086/187460}

\bibitem[{{Mihos} \& {Hernquist}(1996)}]{Mihos&Hernquist1996}
---. 1996, \apj, 464, 641, \dodoi{10.1086/177353}

\bibitem[{{Moreno} {et~al.}(2015){Moreno}, {Torrey}, {Ellison}, {Patton},
  {Bluck}, {Bansal}, \& {Hernquist}}]{Moreno2015}
{Moreno}, J., {Torrey}, P., {Ellison}, S.~L., {et~al.} 2015, \mnras, 448, 1107,
  \dodoi{10.1093/mnras/stv094}

\bibitem[{{Moreno} {et~al.}(2021){Moreno}, {Torrey}, {Ellison}, {Patton},
  {Bottrell}, {Bluck}, {Hani}, {Hayward}, {Bullock}, {Hopkins}, \&
  {Hernquist}}]{Moreno2021}
---. 2021, \mnras, 503, 3113, \dodoi{10.1093/mnras/staa2952}

\bibitem[{{Moreno} {et~al.}(2022){Moreno}, {Danieli}, {Bullock}, {Feldmann},
  {Hopkins}, {{\c{c}}atmabacak}, {Gurvich}, {Lazar}, {Klein}, {Hummels},
  {Hafen}, {Mercado}, {Yu}, {Jiang}, {Wheeler}, {Wetzel},
  {Angl{\'e}s-Alc{\'a}zar}, {Boylan-Kolchin}, {Quataert},
  {Faucher-Gigu{\`e}re}, \& {Kere{\v{s}}}}]{Moreno2022}
{Moreno}, J., {Danieli}, S., {Bullock}, J.~S., {et~al.} 2022, Nature Astronomy,
  6, 496, \dodoi{10.1038/s41550-021-01598-4}

\bibitem[{{Muratov} {et~al.}(2015){Muratov}, {Kere{\v{s}}},
  {Faucher-Gigu{\`e}re}, {Hopkins}, {Quataert}, \& {Murray}}]{Muratov2015}
{Muratov}, A.~L., {Kere{\v{s}}}, D., {Faucher-Gigu{\`e}re}, C.-A., {et~al.}
  2015, \mnras, 454, 2691, \dodoi{10.1093/mnras/stv2126}

\bibitem[{{Nedelchev} {et~al.}(2019){Nedelchev}, {Coccato}, {Corsini}, {Sarzi},
  {de Zeeuw}, {Pizzella}, {Dalla Bont{\`a}}, {Iodice}, \&
  {Morelli}}]{Nedelchev2019}
{Nedelchev}, B., {Coccato}, L., {Corsini}, E.~M., {et~al.} 2019, \aap, 623,
  A87, \dodoi{10.1051/0004-6361/201832840}

\bibitem[{{Osman} \& {Bekki}(2017)}]{Osman&Bekki2017}
{Osman}, O., \& {Bekki}, K. 2017, \mnras, 471, L87,
  \dodoi{10.1093/mnrasl/slx104}

\bibitem[{{Pandya} {et~al.}(2021){Pandya}, {Fielding},
  {Angl{\'e}s-Alc{\'a}zar}, {Somerville}, {Bryan}, {Hayward}, {Stern}, {Kim},
  {Quataert}, {Forbes}, {Faucher-Gigu{\`e}re}, {Feldmann}, {Hafen}, {Hopkins},
  {Kere{\v{s}}}, {Murray}, \& {Wetzel}}]{Pandya2021}
{Pandya}, V., {Fielding}, D.~B., {Angl{\'e}s-Alc{\'a}zar}, D., {et~al.} 2021,
  \mnras, 508, 2979, \dodoi{10.1093/mnras/stab2714}

\bibitem[{{Peebles}(1969)}]{Peebles1969}
{Peebles}, P.~J.~E. 1969, \apj, 155, 393, \dodoi{10.1086/149876}

\bibitem[{{Pizzella} {et~al.}(2004){Pizzella}, {Corsini}, {Vega Beltr{\'a}n},
  \& {Bertola}}]{Pizzella2004}
{Pizzella}, A., {Corsini}, E.~M., {Vega Beltr{\'a}n}, J.~C., \& {Bertola}, F.
  2004, \aap, 424, 447, \dodoi{10.1051/0004-6361:20047183}

\bibitem[{{Pizzella} {et~al.}(2018){Pizzella}, {Morelli}, {Coccato}, {Corsini},
  {Dalla Bont{\`a}}, {Fabricius}, \& {Saglia}}]{Pizzella2018}
{Pizzella}, A., {Morelli}, L., {Coccato}, L., {et~al.} 2018, \aap, 616, A22,
  \dodoi{10.1051/0004-6361/201731712}

\bibitem[{{Pizzella} {et~al.}(2014){Pizzella}, {Morelli}, {Corsini}, {Dalla
  Bont{\`a}}, {Coccato}, \& {Sanjana}}]{Pizzella2014}
{Pizzella}, A., {Morelli}, L., {Corsini}, E.~M., {et~al.} 2014, \aap, 570, A79,
  \dodoi{10.1051/0004-6361/201424746}

\bibitem[{Price {et~al.}(2017)Price, Kriek, Feldmann, Quataert, Hopkins,
  Faucher-Gigu{\`{e}}re, Kere{\v{s}}, \& Barro}]{Price2017}
Price, S.~H., Kriek, M., Feldmann, R., {et~al.} 2017, ApJ, 844, L6,
  \dodoi{10.3847/2041-8213/aa7d4b}

\bibitem[{{Proshina} {et~al.}(2020){Proshina}, {Sil'chenko}, \&
  {Moiseev}}]{Proshina2020}
{Proshina}, I., {Sil'chenko}, O., \& {Moiseev}, A. 2020, \aap, 634, A102,
  \dodoi{10.1051/0004-6361/201936912}

\bibitem[{{Puerari} \& {Pfenniger}(2001)}]{Puerari&Pfenniger2001}
{Puerari}, I., \& {Pfenniger}, D. 2001, \apss, 276, 909,
  \dodoi{10.1023/A:1017581325673}

\bibitem[{{Raimundo} {et~al.}(2023){Raimundo}, {Malkan}, \&
  {Vestergaard}}]{Raimundo2023}
{Raimundo}, S.~I., {Malkan}, M., \& {Vestergaard}, M. 2023, Nature Astronomy,
  7, 463, \dodoi{10.1038/s41550-022-01880-z}

\bibitem[{{Renaud} {et~al.}(2019){Renaud}, {Bournaud}, {Agertz}, {Kraljic},
  {Schinnerer}, {Bolatto}, {Daddi}, \& {Hughes}}]{Renaud2019}
{Renaud}, F., {Bournaud}, F., {Agertz}, O., {et~al.} 2019, \aap, 625, A65,
  \dodoi{10.1051/0004-6361/201935222}

\bibitem[{{Renaud} {et~al.}(2014){Renaud}, {Bournaud}, {Kraljic}, \&
  {Duc}}]{Renaud2014}
{Renaud}, F., {Bournaud}, F., {Kraljic}, K., \& {Duc}, P.~A. 2014, \mnras, 442,
  L33, \dodoi{10.1093/mnrasl/slu050}

\bibitem[{{Rohr} {et~al.}(2022){Rohr}, {Feldmann}, {Bullock},
  {{\c{C}}atmabacak}, {Boylan-Kolchin}, {Faucher-Gigu{\`e}re}, {Kere{\v{s}}},
  {Liang}, {Moreno}, \& {Wetzel}}]{Rohr2022}
{Rohr}, E., {Feldmann}, R., {Bullock}, J.~S., {et~al.} 2022, \mnras, 510, 3967,
  \dodoi{10.1093/mnras/stab3625}

\bibitem[{{Rubin} {et~al.}(1992){Rubin}, {Graham}, \& {Kenney}}]{Rubin1992}
{Rubin}, V.~C., {Graham}, J.~A., \& {Kenney}, J. D.~P. 1992, \apjl, 394, L9,
  \dodoi{10.1086/186460}

\bibitem[{{Saburova} {et~al.}(2018){Saburova}, {Chilingarian}, {Katkov},
  {Egorov}, {Kasparova}, {Khoperskov}, {Uklein}, \& {Vozyakova}}]{Saburova2018}
{Saburova}, A.~S., {Chilingarian}, I.~V., {Katkov}, I.~Y., {et~al.} 2018,
  \mnras, 481, 3534, \dodoi{10.1093/mnras/sty2519}

\bibitem[{{Segovia Otero} {et~al.}(2022){Segovia Otero}, {Renaud}, \&
  {Agertz}}]{SegoviaOtero2022}
{Segovia Otero}, {\'A}., {Renaud}, F., \& {Agertz}, O. 2022, arXiv e-prints,
  arXiv:2206.08379.
\newblock \doarXiv{2206.08379}

\bibitem[{{Sil'chenko} {et~al.}(2009){Sil'chenko}, {Moiseev}, \&
  {Afanasiev}}]{Silchenko2009}
{Sil'chenko}, O.~K., {Moiseev}, A.~V., \& {Afanasiev}, V.~L. 2009, \apj, 694,
  1550, \dodoi{10.1088/0004-637X/694/2/1550}

\bibitem[{{Sil'chenko} {et~al.}(2019){Sil'chenko}, {Moiseev}, \&
  {Egorov}}]{Silchenko2019}
{Sil'chenko}, O.~K., {Moiseev}, A.~V., \& {Egorov}, O.~V. 2019, \apjs, 244, 6,
  \dodoi{10.3847/1538-4365/ab3415}

\bibitem[{{Somerville} \& {Dav{\'e}}(2015)}]{Somerville&Dave2015}
{Somerville}, R.~S., \& {Dav{\'e}}, R. 2015, \araa, 53, 51,
  \dodoi{10.1146/annurev-astro-082812-140951}

\bibitem[{Springel(2005)}]{Springel2005b}
Springel, V. 2005, {The cosmological simulation code GADGET-2},
  \dodoi{10.1111/j.1365-2966.2005.09655.x}

\bibitem[{{Starkenburg} {et~al.}(2019){Starkenburg}, {Sales}, {Genel},
  {Manzano-King}, {Canalizo}, \& {Hernquist}}]{Starkenburg2019}
{Starkenburg}, T.~K., {Sales}, L.~V., {Genel}, S., {et~al.} 2019, \apj, 878,
  143, \dodoi{10.3847/1538-4357/ab2128}

\bibitem[{{Tacchella} {et~al.}(2016){Tacchella}, {Dekel}, {Carollo},
  {Ceverino}, {DeGraf}, {Lapiner}, {Mandelker}, \& {Primack
  Joel}}]{Tacchella2016a}
{Tacchella}, S., {Dekel}, A., {Carollo}, C.~M., {et~al.} 2016, \mnras, 457,
  2790, \dodoi{10.1093/mnras/stw131}

\bibitem[{{Thakar} \& {Ryden}(1998)}]{Thakar&Ryden1998}
{Thakar}, A.~R., \& {Ryden}, B.~S. 1998, \apj, 506, 93, \dodoi{10.1086/306223}

\bibitem[{{Thakar} {et~al.}(1997){Thakar}, {Ryden}, {Jore}, \&
  {Broeils}}]{Thakar1997}
{Thakar}, A.~R., {Ryden}, B.~S., {Jore}, K.~P., \& {Broeils}, A.~H. 1997, \apj,
  479, 702, \dodoi{10.1086/303915}

\bibitem[{{Tohline} {et~al.}(1982){Tohline}, {Simonson}, \&
  {Caldwell}}]{Tohline1982}
{Tohline}, J.~E., {Simonson}, G.~F., \& {Caldwell}, N. 1982, \apj, 252, 92,
  \dodoi{10.1086/159537}

\bibitem[{{Velliscig} {et~al.}(2015){Velliscig}, {Cacciato}, {Schaye}, {Crain},
  {Bower}, {van Daalen}, {Dalla Vecchia}, {Frenk}, {Furlong}, {McCarthy},
  {Schaller}, \& {Theuns}}]{Velliscig2015}
{Velliscig}, M., {Cacciato}, M., {Schaye}, J., {et~al.} 2015, \mnras, 453, 721,
  \dodoi{10.1093/mnras/stv1690}

\bibitem[{Wetzel {et~al.}(2016)Wetzel, Hopkins, Kim, Faucher-Giguere, Keres, \&
  Quataert}]{Wetzel2016}
Wetzel, A.~R., Hopkins, P.~F., Kim, J.-h., {et~al.} 2016, The Astrophysical
  Journal, \dodoi{10.3847/2041-8205/827/2/L23}

\bibitem[{{Yu} {et~al.}(2021){Yu}, {Bullock}, {Klein}, {Stern}, {Wetzel}, {Ma},
  {Moreno}, {Hafen}, {Gurvich}, {Hopkins}, {Kere{\v{s}}},
  {Faucher-Gigu{\`e}re}, {Feldmann}, \& {Quataert}}]{Yu2021}
{Yu}, S., {Bullock}, J.~S., {Klein}, C., {et~al.} 2021, \mnras, 505, 889,
  \dodoi{10.1093/mnras/stab1339}

\bibitem[{{Yu} {et~al.}(2023){Yu}, {Bullock}, {Gurvich}, {Hafen}, {Stern},
  {Boylan-Kolchin}, {Faucher-Gigu{\`e}re}, {Wetzel}, {Hopkins}, \&
  {Moreno}}]{Yu2023}
{Yu}, S., {Bullock}, J.~S., {Gurvich}, A.~B., {et~al.} 2023, \mnras, 523, 6220,
  \dodoi{10.1093/mnras/stad1806}

\bibitem[{{Zhou} {et~al.}(2022){Zhou}, {Chen}, {Shi}, {Bizyaev}, {Guo}, {Bao},
  {Xu}, {Yu}, \& {Brownstein}}]{Zhou2022}
{Zhou}, Y., {Chen}, Y., {Shi}, Y., {et~al.} 2022, \mnras, 515, 5081,
  \dodoi{10.1093/mnras/stac2016}

\bibitem[{{Zinchenko}(2023)}]{Zinchenko2023}
{Zinchenko}, I.~A. 2023, \aap, 674, L7, \dodoi{10.1051/0004-6361/202346846}

\bibitem[{Zolotov {et~al.}(2015)Zolotov, Dekel, Mandelker, Tweed, Inoue,
  DeGraf, Ceverino, Primack, Barro, \& Faber}]{Zolotov2015}
Zolotov, A., Dekel, A., Mandelker, N., {et~al.} 2015, MNRAS, 450, 2327,
  \dodoi{10.1093/MNRAS/STV740}

\end{thebibliography}
\bibliographystyle{aasjournal}



\end{document}